\documentclass[prb,preprint]{revtex4-1}

\usepackage{amsmath}
\usepackage[dvips]{psfrag,graphicx}
\usepackage{color}
\usepackage{amssymb}
\usepackage{bbold}
\usepackage{bm}
\begin{document}
\title{Arbitrary qubit transformations on tuneable Rashba rings}

\author{A.  Kregar}

\affiliation{J. Stefan Institute, Ljubljana, Slovenia}

\author{J. H. Jefferson}
\affiliation{Department of Physics, Lancaster University, Lancaster LA1 4YB, UK}

\author{A. Ram\v sak}
\affiliation{J. Stefan Institute, Ljubljana, Slovenia}
\affiliation{Faculty of Mathematics and Physics, University of Ljubljana, Ljubljana, Slovenia}

\date{\today}
\begin{abstract}
An exact solution is presented for the time-dependent wavefunction of a Kramers doublet which propagates around a quantum ring with tuneable Rashba spin-orbit interaction.  By propagating in segments it is shown that Kramers-doublet qubits may be defined for which transformations on the Bloch sphere may be performed for an integral number of revolutions around the ring. The conditions for full coverage of the Bloch sphere are determined and explained in terms of sequential qubit rotations due to electron motion along the segments, with change of rotation axes between segments due to adiabatic changes in the Rashba spin-orbit interaction. Prospects and challenges for possible realizations are discussed for which rings based on InAs quantum wires are promising candidates.
\end{abstract}

\maketitle
\section{Introduction}
In last few decades, a promising new branch of electronics, called spintronics, has emerged. In contrast to ordinary electronics, where the electron charge is used as a fundamental resource, spintronics use electron spin, which gives improved performance due to longer coherence times and lower power consumption \cite{Wolf2001,Zutic2004,Rashba2007}. Spintronic systems are also believed to be one of the most suitable candidates for the realization of quantum computers with spin states being used as qubits \cite{Awschalom2013}. The use of spintronics is still somewhat limited since magnetic fields, which can be used to manipulate electron spin, are hard to control on small scales \cite{Zutic2004}. 

One possible solution to this problem might be the use of spin-orbit interaction (SOI) \cite{Winkler2003,Engel2007} in semiconductor heterostructures, a system in which future spintronic devices will most likely be fabricated \cite{Zutic2004}. Two types of SOI are present in such heterostructures: Dresselhaus type (DSOI) \cite{Dresselhaus1955} of interaction emerges due to bulk inversion asymmetry of a crystal while the Rashba type spin-orbit interaction (RSOI) \cite{Rashba1960}  is a consequence of structural inversion asymmetry of the potential, confining the two-dimensional electron gas (2DEG). RSOI is especially promising for use in spintronic devices since its strength can be tuned externally using voltage gates. \cite{Nitta1997,Schapers1998} Following the proposal of using spin-orbit coupling in field effect transistors in 1990 \cite{Datta1990}, many ideas of using SOI in new spintronic devices emerged, usually based on two-dimensional structures, fabricated in semiconductor heterostructures. \cite{Nitta1999,Wolf2001,Schliemann2003,Zutic2004,Wunderlich2010,Stajic2013}

Quantum rings are often used to describe such devices due to their geometric simplicity which nevertheless belies rich mesoscopic behaviour. In the presence of a magnetic field, quantum rings exhibit some interesting phenomena, such as the well known Aharonov-Bohm effect and persistent currents \cite{Buttiker1983,Fuhrer2001}, which have very similar counterparts even in the absence of externally applied magnetic fields \cite{Aronov1993,Qian1994,Malshukov1999}. It was shown quite early on that SOI on a ring can be seen as an effective spin dependent flux \cite{Meir1989}, similar to magnetic flux through a ring, leading to an electrical counterpart of Aharonov-Bohm effect, the so called Aharonov-Casher effect \cite{Aharonov1984,Konig2006}. 

After the correct form of the Hamiltonian for electrons on a ring in presence of SOI had been derived \cite{Meijer2002}, a vast literature emerged studying its properties. Transmission of electrons through a ring with two \cite{Aronov1993,Nitta1999,Peeters2004,Frustaglia2004,Aeberhard2005,Wang2005} or more \cite{Souma2004,Lozano2005,Shelykh2005,Harmer2006} attached leads has been studied in presence of an external magnetic field \cite{Frustaglia2004,Peeters2004,Lozano2005,Wang2005}, showing that rings can be used as spin dependent interferometers, enabling spin filtering \cite{Souma2004,Frustaglia2004,Shelykh2005,Lobos2008,Bagraev2010}  as well as measurement \cite{Morpurgo1998,Konig2006,Maiti2014} of the strenght of spin-orbit coupling. Spin dependent conductance could also be used to perform single qubit transformations on electrons travelling through the ring \cite{Foldi2005}. Electron states on a ring with both RSOI and DSOI have also been studied, showing the presence of spin-dependent persistent currents and non-trivial spin polarization, both depending on SOI strength \cite{Splettstoesser2003,Sheng2006,Yang2006,Chen2007,Nowak2009}.

The motion of an electron through the crystal, controlled by an external confining potential, enables controlled spin rotation for which analytical results have been obtained in the adiabatic limit \cite{San-Jose2008,Golovach2010}. For the special case of a linear wire, analytical results have also been obtained for arbitrary fast driving in one dimension \cite{Cadez2014} though, due to the fixed axis of rotation, rotations on the Bloch sphere are severely restricted.  In this paper, we show that this limitation can be overcome when the electron is constrained to move around a closed ring instead of a linear wire. Exact analytical solutions for practical limiting cases are given and it is proved that performing a general single-qubit transformation is possible simply by moving the electron around the ring using an external driving potential in the presence of the Rashba type SOI. 

In Sec. \ref{sec:model} we present a model Hamiltonian for an electron in a mesoscopic ring with Rashba coupling and confined in a gate-induced moving potential well. Using unitary transformations, solutions of the time-dependent  Schr\"odinger equation are then derived in Sec. \ref{sec:unitrans} for (A) the electron moving through a segment of the ring, giving rise to a spin-rotation, and (B) adiabatic change in the SOI strength with the potential well held fixed, which changes the axis of rotation. These evolutions are combined in Sec. \ref{sec:comb_evo} to give solutions for which the electron propagates around the ring in segments between which the SOI strength is changed adiabatically, resulting in cumulative spin and axis rotations. Qubits are then defined in Sec. \ref{sec:qtrans} as superpositions of the Kramers ground-doublet components at some fixed position on the ring. We then show that qubit transformations may be performed for an integral number of rotations of a (Gaussian) wave-packet around the ring and discuss allowed coverage of the associated Bloch sphere, including conditions for complete coverage. We conclude in Sec. \ref{sec:discussion} with a discussion of future prospects and challenges for possible physical realizations, using estimates based on current experimental results.

\section{Model}
\label{sec:model}

\begin{figure}[htbp]
\includegraphics[scale=1]{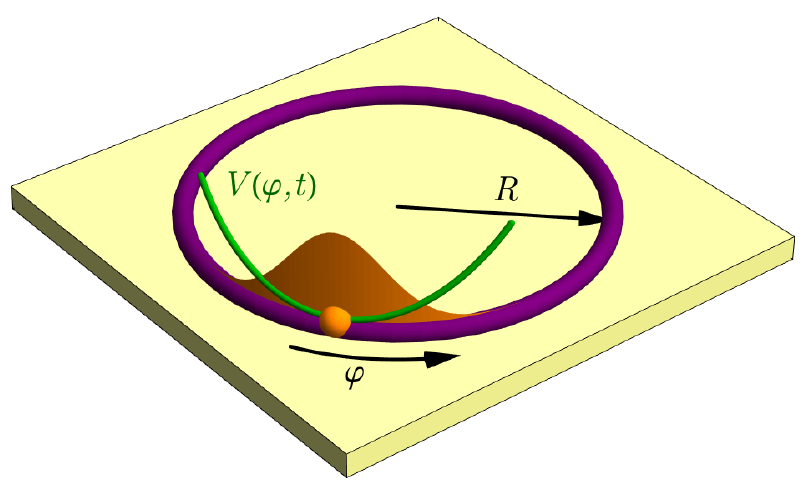}
\caption{Schematic presentation of the system. The position of the electron (orange), confined by potential well $V$ (green) on a ring (purple) of radius $R$, is described by coordinate $\varphi$.}
\label{fig:fig1}
\end{figure}

We consider an electron, confined in a narrow ring with the Rashba coupling and described by the Hamiltonian \cite{Meijer2002} 
\begin{equation}
\label{eq:Hstart}
H = \epsilon p_\varphi^2 + \epsilon \alpha(t) \left( \sigma_\rho p_\varphi - \frac{i}{2} \sigma_\varphi \right) + V(\varphi,t),
\end{equation}
where
\begin{equation}
\label{eq:def1}
\epsilon \equiv \frac{\hbar^2}{2 m R^2}, \qquad \alpha(t) \equiv \frac{2 m R \alpha_{R}(t)}{\hbar},
\end{equation}
with $R$ being the ring radius, $m$ the electron effective mass and $\alpha_R$ the Rashba coupling. $V$ is a potential well, which confines and moves the electron around the ring as shown schematically in Fig.~\ref{fig:fig1}.

The position of the electron is described by the  angle $\varphi$. Only scaled momentum along the ring is relevant, $ p_\varphi = -i \frac{\partial}{\partial \varphi}$, with the commutation relation
\begin{equation}
\left[ \varphi , p_\varphi \right] = i,
\end{equation}
and the spin operators in the cylindrical coordinate system,
\begin{equation}
\bm{\sigma} =  \sigma_\rho {\bf \hat{e}}_\rho + \sigma_\varphi  {\bf \hat{e}}_\varphi +\sigma_z  {\bf \hat{e}}_z, 
\end{equation}
are given by
\begin{eqnarray}
\sigma_\rho\left( \varphi \right) &=\phantom{-} & \sigma_x \cos \varphi + \sigma_y \sin \varphi, \\
\sigma_\varphi\left( \varphi \right) &=  - &\sigma_x \sin \varphi + \sigma_y \cos \varphi,
\end{eqnarray}
with  ${\bf \hat{e}}_\rho,  {\bf \hat{e}}_\varphi , {\bf \hat{e}}_z$ being orthogonal unity vectors.

The Rashba coupling $\alpha(t)$ can be controlled by the application of an external electric field perpendicular to the 2DEG plane. The amplification of Rashba coupling, given as a ratio between amplified value and intrinsic value $k=\alpha_{amp}/\alpha_{int}$, is up to about $1.5$ for electrons in a 2DEG \cite{Nitta1997} and up to $k = 6$ for electrons on an InAs nanowire \cite{Liang2012}. 

The potential energy $V(\varphi,t)$ is a periodic function with period $2 \pi$ and minimum at $\xi(t)$. Here we consider cases which can be expanded around the minimum and accurately described by the harmonic form,
\begin{equation}
V(\varphi, t) = \omega^2 \left[ \varphi - \xi(t) \right]^2
\end{equation}
where the constant of proportionality, $\omega^2 = \frac{\partial^2 V}{d \varphi^2}\big|_{\varphi=\xi(t)}$, depends on the detail of the applied gate.

\section{Unitary transformations}
\label{sec:unitrans}

 The quantities $\alpha(t)$ and $\xi(t)$ are determined by the external electric field and will be the main parameters of the qubit transformation. The time evolution of the wavefunction is driven by the time-dependent Hamiltonian Eq.~\eqref{eq:Hstart}, giving the time-dependent Schr\"odinger equation
\begin{equation}
\label{eq:sch0}
i \hbar \frac{\partial \psi (t)}{\partial t}=H(t) \psi(t).
\end{equation}

The equation can be solved analytically using unitary transformations. For a time-dependent transformation $U(t)$, we operate on Eq.~\eqref{eq:sch0} with $U(t)$ and the Schr\"odinger equation becomes
\begin{equation}
\label{eq:sch1}
i \hbar \frac{\partial \psi' (t)}{\partial t}=H'(t) \psi'(t)   
\end{equation}
with transformed Hamiltonian
\begin{equation}
\label{eq:Ht1}
H'(t) = U(t) H(t) U^\dagger(t) - i \hbar U(t) \dot{U}^\dagger(t)
\end{equation}
and corresponding wavefunction
\begin{equation}
\psi'(t) = U(t) \psi(t).
\end{equation}

We notice that $\sigma_\rho$ and $\sigma_\varphi$, which both depend on $\varphi$, can be transformed to the form of standard Pauli matrices $\sigma_x$ and $\sigma_y$ by the position dependent unitary transformation
\begin{equation}
U_z = \exp \left(i \frac{\varphi}{2} \sigma_z \right).
\end{equation}
Since $U_z$ is independent of time, the second term in the transformed Hamiltonian Eq.~\eqref{eq:Ht1} is zero and, using $U_z \sigma_{\rho,\varphi} U_z^\dagger = \sigma_{x,y}$, we get
\begin{align}
\label{eq:H1}
H'(t) &= U_z H(t) U_z^\dagger = \\ \nonumber
& = \epsilon p_\varphi^2 + \epsilon p_\varphi {\bm \alpha} (t) \cdot {\bm \sigma} + \omega^2 \left[\varphi - \xi(t) \right]^2 + \frac{\epsilon}{4}.
\end{align}
The kinetic part $p_\varphi^2$ and the external potential are diagonal in spin and they do not change whilst the unnormalized vector
\begin{equation}
{\bm \alpha}(t) =\left( \alpha(t),0,-1 \right)
\end{equation}
 becomes the axis of spin rotation due to SOI.
 
In order to solve the Schrodinger equation for the Hamiltonian Eq.~\eqref{eq:H1}, we shall restrict ourselves to two types of time-dependence. In the first we drive the electron around the ring by changing the position of the potential minimum, $\xi(t)$, whilst keeping the Rashba coupling constant. Alternatively, we keep the position of the minimum fixed whilst varying the Rashba coupling.

\subsection{Time-dependent driving potential}
\label{subsec:t_pot}
If $\alpha$ is kept constant, the SO part of Hamiltonian Eq.~\eqref{eq:H1} can be eliminated using the further transformation
\begin{equation}
\label{eq:Ualpha}
U_\alpha = \exp \left( i \frac{\varphi}{2} {\bm \alpha}\cdot {\bm \sigma}  \right).
\end{equation}
resulting in the transformed Hamiltonian
\begin{align}
\label{eq:H2}
H''(t) &= U_\alpha H'(t) U_\alpha^\dagger =\nonumber \\
 &=\epsilon p_\varphi^2 + \omega^2\left[\varphi - \xi(t) \right]^2 - \frac{\epsilon \alpha^2 }{4}.
\end{align}
This Hamiltonian is exactly soluble for any driving $\xi(t)$ \cite{Cadez2014}. We use the transformation
\begin{equation}
\label{eq:trans_xi}
U^\dagger_\xi(t) = e^{i \phi_{0}(t)} e^{i \frac{\hbar}{2 \epsilon} \dot{\varphi}_c(t) \varphi} e^{-i \varphi_c(t) p_\varphi},
\end{equation}
where $\varphi_c(t)$ is the classical harmonic oscillator solution, driven by $\xi(t)$
\begin{equation}
\ddot{\varphi}_c(t) + \Omega^2 \varphi_c(t) = \Omega^2 \xi(t)
\end{equation}
and $\Omega = \frac{\omega}{R} \sqrt{\frac{2}{m}}$. The first term in the transformation Eq.~\eqref{eq:trans_xi} is a time-dependent phase factor independent of spin
\begin{equation}
\phi_{0}(t) = \int_0^t L_\xi(t') dt'  -\frac{\hbar}{2 \epsilon} \varphi_c(t') \dot{\varphi}_c(t') \big|_0^t 
\end{equation}
containing a time integral of the Lagrangian
\begin{equation}
L_\xi(t) = \frac{\hbar}{4 \epsilon} \left(\dot{\varphi}_c^2(t)- \Omega^2 \left[\varphi_c(t) - \xi(t)\right]^2 \right).
\end{equation}

The second term of the transformation depends on $\varphi$ and gives additional momentum to the electron, proportional to the classical velocity $\dot{\varphi}_c$. The third term is the translation operator, which shifts the wave function by $\varphi_c(t)$. The transformation removes time dependence of the potential,
\begin{align}
\label{eq:H3}
H'''(t) &= U_\xi H''(t) U_\xi^\dagger =  \epsilon p^2 + \omega^2 \varphi^2 -  \frac{\epsilon \alpha^2 }{4}.
\end{align}
The resulting time-independent Hamiltonian is that of a harmonic oscillator with eigenstates
\begin{equation}
\psi''' = g_{n} (\varphi) = \frac{1}{\sqrt{2^n n! \sigma \sqrt{\pi}}} e^{-\frac{\varphi^2}{2 \sigma^2}} H_n\left( \frac{\varphi}{\sigma} \right)
\end{equation}
with $\sigma = \left( \frac{\epsilon}{\omega^2} \right)^{1/4}$, $H_n$ being Hermite polynomials, and eigenenergies $E_{n} = \hbar \Omega \left( n +\frac{1}{2} \right) - \epsilon \alpha^2 /4$. In accord with the quadratic expansion of potential well function we assume also that confinement is sufficiently strong that the tails of the function are negligible at $\varphi = \pm \pi$.

It is important to note that in this limit, the energy does not depend on the spin of the electron, so we have two degenerate states with different spin, forming the Kramers doublet state. We focus here on the ground-state doublet $n=0$, for which the time-dependent solution of Schr\" odinger's equation is given by
\begin{align}
\label{eq:psi_t1}
\psi(\varphi,t) =& U_z^\dagger(\varphi) U_\alpha^\dagger(\varphi) U_\xi(t)^\dagger \psi'''(\varphi) = \nonumber \\ 
=&  e^{-i \frac{E_{n}}{\hbar} t} e^{i \phi_{0}(t)} e^{i\frac{\hbar}{2 \epsilon} \dot{\varphi}_c(t) \varphi} U^\dagger_z(\varphi) U_{\alpha}^\dagger(\varphi) g_0(\varphi - \varphi_c(t)) \chi_i.
\end{align}
The initial state of the system is described by the initial position $\varphi_c(t)$ of the wavepacket $g_0$, i.e. $\varphi_c(0)$, its initial momentum $\dot{\varphi}_c(0)$, and the spinor $\chi_i = c_{i, \uparrow} \chi_\uparrow + c_{i ,\downarrow} \chi_\downarrow$. The time dependent phase, irrelevant for spin transformations, will be labelled as $\phi_{\xi}(t) = \phi_{0}(t) - \frac{E_{n}}{\hbar} t$ in the rest of the paper.

\subsection{Time-dependent Rashba coupling}
\label{subsec:a_pot}
To solve the Schr\"odinger equation when changing the Rashba coupling $\alpha(t)$, with constant driving, $\xi(t) = \xi_0$, we first transform the Hamiltonian Eq.~\eqref{eq:H1} with the time independent displacement transformation $U_\xi = e^{-i \xi_0 p_\varphi}$, resulting in
\begin{align}
H''(t) &= U_\xi H'(t) U_\xi^\dagger =\nonumber \\
 &=\epsilon p_\varphi^2 + \epsilon p_\varphi {\bm \alpha} (t) \cdot {\bm \sigma} + \omega^2 \varphi ^2 + \frac{\epsilon}{4}.
\end{align}
Since the direction of the spin rotation axis, ${\bm \alpha}(t)$, depends on $\alpha$, the time-dependent spin-orbit part of the Hamiltonian can only be eliminated analytically in the limit of adiabatically changing $\alpha(t)$. In this limit, the term $- i \hbar U_\alpha(t) \dot{U}_\alpha^\dagger(t)  \propto \dot{\alpha}(t) \ll 1$ will be neglected, resulting in the transformed Hamiltonian
\begin{align}
H'''(t) &= U_\alpha(t) H'(t) U_\alpha(t)^\dagger =\nonumber \\
 &=\epsilon p_\varphi^2 + \omega^2 \varphi ^2 + \frac{\epsilon \alpha(t)^2}{4 },
\end{align}
with solutions in the time-dependent two-dimensional ground Hilbert subspace spanned by states
\begin{equation}
\label{eq:psi_st0}
\psi_s(\varphi,t) =  U^\dagger_z(\varphi) U_{\alpha(t)}^\dagger(\varphi - \xi_0) g_0(\varphi - \xi_0) \chi_s.
\end{equation}
As we show in the Appendix~\ref{app:appA}, the actual state at time $t$ is
\begin{equation}
\psi(\varphi,t) = e^{i \phi_{\alpha}(t)}g_0(\varphi-\xi_0) U_z^\dagger(\varphi) U_\alpha^\dagger(\varphi-\xi_0) U_y^\dagger(\tilde{\vartheta}_{\alpha(t)}) \chi_i,
\end{equation}
where $U_y^\dagger(\tilde{\vartheta}_{\alpha(t)}) = e^{-i \tilde{\vartheta}_{\alpha(t)} \sigma_y}$ is as small rotation around $y$-axis which depends on $\alpha(t)$ and the spread ($\sigma$) of the ground state oscillator wavefunction $g_0$.

The spinor $\chi_i = c_{0,\uparrow} \chi_\uparrow + c_{0,\downarrow} \chi_\downarrow$ describes the initial spin state at $t=0$ and we note that $\psi(\varphi,t)$ may therefore be written in the equivalent form
\begin{equation}
\label{eq:psit_alpha}
\psi(\varphi,t) =  e^{i \phi_{\alpha}(t)} \sum_s \psi_{\alpha(t),\xi_0,s}(\varphi) c_{0 s},
\end{equation}
where 
\begin{equation}
\label{eq:Kramers}
\psi_{\alpha(t),\xi_0,s}(\varphi) =g_0(\varphi-\xi_0) U_z^\dagger(\varphi) U_\alpha^\dagger(\varphi-\xi_0) U_y^\dagger(\tilde{\vartheta}_{\alpha(t)}) \chi_s
\end{equation}
are appropriate time-dependent Kramers states.

\section{Combined evolutions}
\label{sec:comb_evo}

The general transformation is a sequence of $m$ shifts of potential with change of Rashba coupling after each shift. We choose the electron wave packet before $i$-th shift, i. e. at time $t_{i-1}$, to be centred at $\varphi_c(t_{i-1}) = \varphi_{i-1}$ with Rashba coupling $\alpha_{i-1}$. We also choose the wavefunction to be stationary, i. e. $\dot{\varphi}_c(t_i)=0$. The state of the system can therefore be written as a superposition of Kramers states
\begin{equation}
\psi(\varphi,t_{i-1}) = \sum_s \psi_{\alpha_{i-1},\varphi_{i-1},s}(\varphi) c_{i-1,s}.
\end{equation}
We then change $\alpha$ adiabatically from $\alpha_{i-1}$ to $\alpha_{i}$ over time interval $\Delta t_{\alpha,i}$. According to Eq.~\eqref{eq:psit_alpha}, this will not change the coefficients $c_{i-1,s}$, resulting in a state:
\begin{equation}
\psi(\varphi,t_{i-1} +\Delta t_{\alpha,i} ) = e^{i \phi_{\alpha,i}} \sum_s \psi_{\alpha_i,\varphi_{i-1},s}(\varphi) c_{i-1,s},
\end{equation}
where $\phi_{\alpha,i} = \int_{t_{i-1}}^{t_{i-1} +\Delta t_{\alpha,i}} \phi_{\alpha}(t') dt'$ is the acquired spin-independent phase.

\begin{figure}[htbp]
\centering
\includegraphics[scale=1]{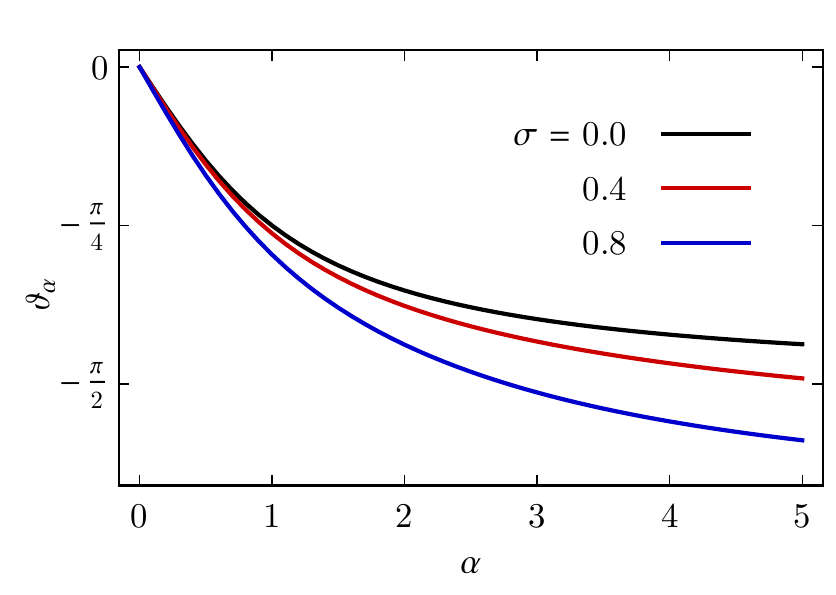}
\caption{Angle $\vartheta_{\alpha}$, plotted as a function of $\alpha$ for 3 values of $\sigma$. For $\sigma=0$ (black), the angle equals $\vartheta_\alpha = -\arctan \alpha$, while for larger $\sigma$ ($0.4$ (red) and $0.8$ (blue)), the angle $\tilde{\vartheta}_\alpha$ is added.}
\label{fig:fig2}
\end{figure}

This transformation is followed by an appropriate shift of potential $\xi(t)$ such that $\varphi_c(t_i) = \varphi_i$ and $\dot{\varphi}_c(t_i)=0$. The shift requires time $\Delta t_{\xi,i}$, resulting in $t_i - t_{i-1} = \Delta t = \Delta t_{\alpha,i} +\Delta t_{\xi,i}$. To determine the effect of this transformation on the Kramers coefficients, we first write the wavefunction in a form, similar to Eq.~\eqref{eq:psi_t1}:
\begin{align}
\psi(\varphi,t_{i-1} +\Delta t_{\alpha,i} ) = e^{i \phi_{\alpha,i}} \sum_s \psi_{\alpha_i,\varphi_{i-1},s}(\varphi) c_{i-1,s} = e^{i \phi_{\alpha,i}} g_0(\varphi - \varphi_{i-1}) U^\dagger_z(\varphi) U_{\alpha_i}^\dagger(\varphi) \chi_{i-1}
\end{align}
The spinor $\chi_{i-1}$ is defined as 
\begin{equation}
\label{eq:c0}
\chi_{i-1} = U_{\alpha_i}^\dagger(-\varphi_{i-1}) U_y^\dagger(\tilde{\vartheta}_{\alpha_{i}}) \sum_{s} \chi_s c_{i-1,s},
\end{equation}
which follows from the definition of Kramers states, Eq.~\eqref{eq:Kramers}. According to Eq.~\eqref{eq:psi_t1}, the prescribed shift of the potential well will only result in the shift of wavepacket $g_0$ together with an additional spin-independent phase:
\begin{align}
\label{eq:psi_ti}
\psi(\varphi,t_i = t_{i-1} +\Delta t) =& e^{i \phi_{\xi,i}} e^{i \phi_{\alpha,i}} g_0(\varphi - \varphi_{i}) U^\dagger_z(\varphi) U_{\alpha_i}^\dagger(\varphi) \chi_{i-1} = \nonumber \\
=& e^{i \phi_i} \sum_s \psi_{\alpha_{i},\varphi_{i},s}(\varphi) c_{i,s}
\end{align}
where $\phi_{\xi,i} = \int_{t_{i-1}}^{t_{i-1} +\Delta t_{\alpha,i}} \phi_{\xi}(t') dt'$ with $\phi_i = \phi_{\alpha,i} + \phi_{\xi,i}$. The new coefficients $c_{i,s}$ are obtained by substituting Eq.~\eqref{eq:c0} into Eq.~\eqref{eq:psi_ti}, enabling the new coefficients to be expressed as a linear combination of the old coefficients,
\begin{equation}
c_{i,s} = \sum_{s'} \chi_s^\dagger U^\dagger_i \chi_{s'} c_{i-1,s'}.
\end{equation}
The spin rotation operator $U_i^\dagger$ depends on $\alpha_i$, $\Delta \varphi_i$ and the width $\sigma$ of the Gaussian envelope $g_0$:
\begin{equation}
U^\dagger_i =  U_y(\tilde{\vartheta}_{\alpha_{i}}) U_{\alpha_i}^\dagger(\Delta \varphi_i) U_y^\dagger(\tilde{\vartheta}_{\alpha_{i}}),
\end{equation}
which can be written in more intuitive form
\begin{equation}
U^\dagger_i = e^{- i \frac{\gamma_i}{2} {\bf n}_{i} \cdot {\bm \sigma}}, \quad {\bf n}_{i} = (\sin  \vartheta_{\alpha_i} , 0, \cos  \vartheta_{\alpha_i}),
\end{equation}
with rotation axis ${\bf n}_{i}$ tilted by $\vartheta_{\alpha_i}$ from $z$ to $x$ direction and rotation angle $\gamma_i$, defined as:
\begin{equation}
 \vartheta_{\alpha_i} = \tilde{\vartheta}_{\alpha_{i}}  - \arctan \alpha_i , \quad \gamma_i = -\Delta \varphi_i \sqrt{1 + \alpha_i^2}.
\end{equation}
The tilting angle $\vartheta_{\alpha_i}$ is plotted in Fig.~\ref{fig:fig2} for various values of $\sigma$.

Using this result, the full transformation, composed of $m$ changes of $\alpha$ and $\xi$, can be obtained as a product of spin rotations $U^\dagger = \prod_{i=1}^m U_i^\dagger $ and summation of spin-independent phases $\phi = \sum_{i=1}^m \phi_i$.

\section{Qubit transformations}
\label{sec:qtrans}

\begin{figure}[htbp]
\includegraphics[scale=1]{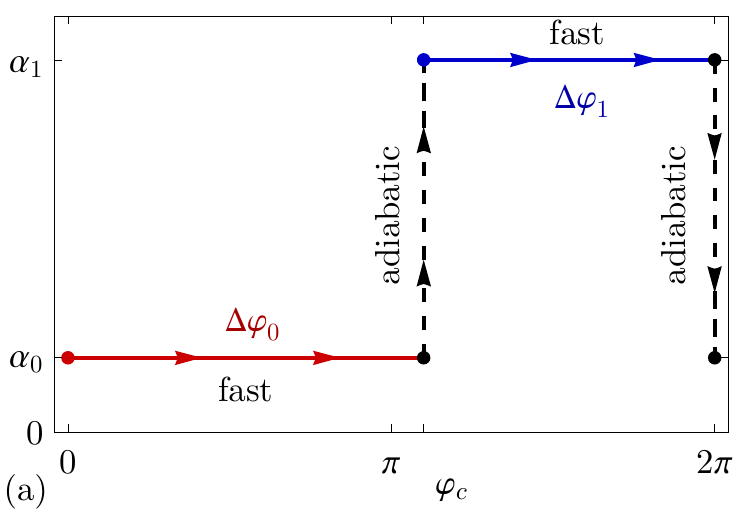}
\includegraphics[scale=1]{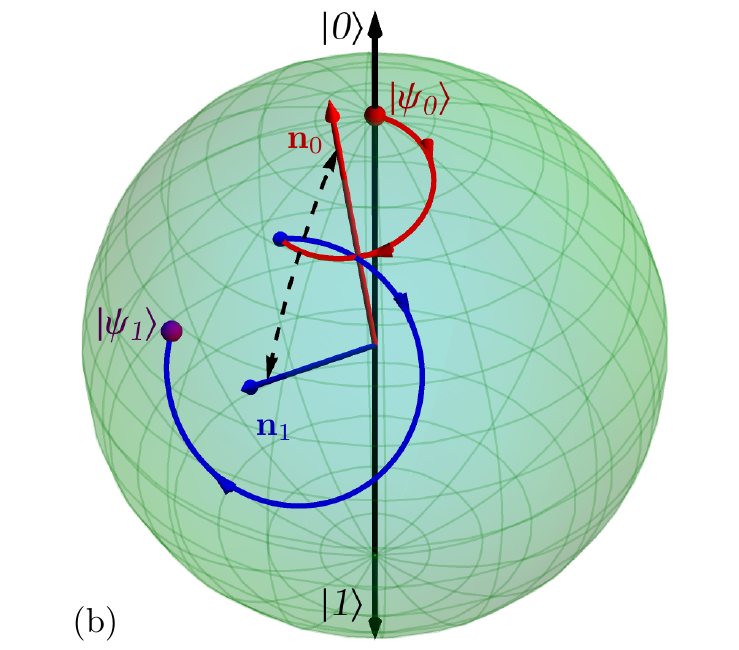}
\caption{A simple example of a qubit transformation with two shifts of the wave function. The movement of electron around the ring (red and blue lines in (a)) causes a rotation of the initial state $\left| \psi_0 \right\rangle= \left| 0 \right\rangle$ on the Bloch sphere (b) around axis ${\bf n}_0$ (red line) or ${\bf n}_1$ (blue line), bringing it to the final state $\left| \psi_1 \right\rangle $. Orientation of the rotation axis is determined by the strength of Rashba coupling $\alpha$, taking the values $\alpha_0 = 0.3$ and $\alpha_1 = 1.5$.}
\label{fig:fig3}
\end{figure}

After a qubit transformation we require that the wave packet returns to its initial position, i.e. $\sum_{i=1}^m \Delta\varphi_i = 2 \pi n$, and also that the Rashba coupling is restored to its initial value, $\alpha_0$. This latter process does not change the coefficients $c_{m, s}$, but returns the system in its initial Kramers subspace, which can now be used to define the qubit basis
\begin{equation}
\left| 0 \right \rangle = \left| \psi_{\alpha_{0},\varphi_{0},\uparrow} \right\rangle, \quad \left| 1 \right \rangle = \left| \psi_{\alpha_{0},\varphi_{0},\downarrow}\right\rangle.
\end{equation}
The initial qubit state, defined by coefficient $c_{0 s}$
\begin{equation}
\left| \psi \right\rangle = \left| 0 \right \rangle c_{0, \uparrow} + \left| 1 \right \rangle c_{0, \downarrow}
\end{equation}
will be transformed into the same qubit basis with different coefficients $c_{m, s}$,  
\begin{equation}
(c_{m, \uparrow}, c_{ m,\downarrow} ) = \left( \cos \frac{\Theta}{2}, e^{i \Phi} \sin \frac{\Theta}{2} \right),
\end{equation}
 representing a point ${\bf r} = \left( \sin \Theta \cos \Phi, \sin \Theta \sin \Phi, \cos \Theta \right) $ on the Bloch sphere, spanned by qubit states $\left| 0 \right \rangle$ and $\left| 1 \right \rangle$. This also means that the spin rotation $U^\dagger$ can be represented by a rotation on the Bloch sphere, composed of intermediate rotations $U_i^\dagger$, determined by shifts of potential and Rashba coupling. Fig.~\ref{fig:fig3} shows a simple example of such a transformation, realized by $m=2$ shifts of confining potential well, starting with qubit state $\left| 0\right\rangle$ and in the high-confinement limit, $\sigma =0 $. First the electron is shifted by an angle $\Delta \varphi_0 = 1.1 \pi$ (red), causing a rotation around the axis ${\bf n}_0$ for an angle $\gamma_0 = - \pi \sqrt{1+\alpha_0^2}$. $\alpha$ is then increased adiabatically to $\alpha_1$ (black dashed arrows) and the electron is then shifted by $\Delta \varphi_1 = 0.9 \pi$ (blue), returning back to its initial position and causing additional rotation around ${\bf n}_1$ for an angle $\gamma_1 = \pi \sqrt{1+\alpha_1^2}$. Finally, $\alpha$ is reduced adiabatically back to $\alpha_0$, returning to a state in the qubit space.

Every rotation $U^\dagger$ can be described by $\Theta$ and $\Phi$ as polar coordinates of the point ${\bf r}$, obtained from ${\bf r}_0 = (0,0,1)$ by the rotation $U^\dagger$. To verify that we are able to achieve arbitrary rotations by the wavefunction transformation, described in Sec.~\ref{sec:comb_evo}, we need to check that all allowed $(\Theta,\Phi)$ pairs may be reached.

We will restrict to simplified cases in which only two values of $\alpha$ are used during the transformation: the intrinsic value $\alpha_{0}$ and an amplified value $\alpha_1 = k \alpha_0$, where $k$ is defined by material properties. For even $i$ in transformations $U_i^\dagger$, the value $\alpha_0$ will be used and for odd $i$ $\alpha_1$ will be used. This results in spin rotations around only two different axes. We argue that the transformation is most efficient when the angle between the axes of rotation $\delta = \vartheta_{\alpha_1} - \vartheta_{\alpha_0}$ is as large as possible. Indeed, for $\delta=\pi/2$, a general rotation can be realized by only one rotation around each axis. In our case, the maximum angle is limited by the amplification factor $k$. In the high-confinement limit ($\sigma=0$), the small contribution $\tilde{\vartheta}_\alpha$ to the tilting angle $\vartheta_{\alpha_i}$ vanishes, so angle $\delta$ is maximal when $\alpha_0 = \frac{1}{\sqrt{k}}$ and $\alpha_1 = \sqrt{k}$. For a given $k$, this choice is made by selecting the ring radius in accord with Eq.~\eqref{eq:def1}: $R = \frac{\hbar}{2 m \sqrt{k}\alpha_{R,0}}$, with intrinsic value of Rashba coupling $\alpha_{R,0}$. These two values of $\alpha$ will be used in our further calculations. The high-confinement limit will also be used in the rest of the paper since finite $\sigma$ does not qualitatively change the results.

To prove that any point on the Bloch sphere can be reached, we first rotate the coordinate system around the $y$ axis so that the new $z'$ axis is pointing in the direction ${\bf n}(\alpha_0)$. The two rotations are now a rotation around the $z'$ axis  (transformation $U_0^\dagger$) and around the axis ${\bf n}_\delta = ( \sin \delta , 0, \cos \delta) $ (transformation $U_1^\dagger$).

\begin{figure}[htbp]
\centering
\includegraphics[scale=1]{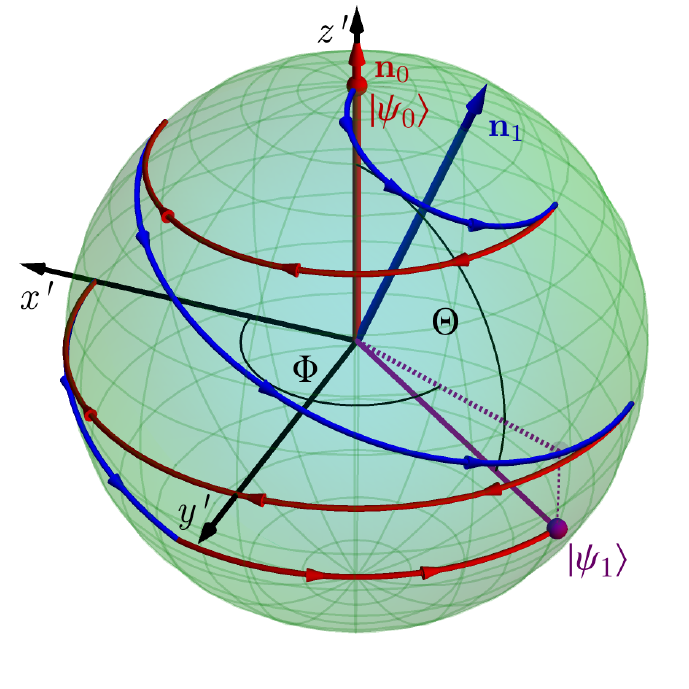}
\caption{Controlled transformation from qubit state $\left| \psi_0 \right\rangle$ (red dot), corresponding to $\Theta'=0$, to the target state $\left| \psi_1 \right\rangle$ (purple dot), corresponding to $\Theta'$ and $\Phi'$. By alternating rotations $U_0^\dagger$ (axis ${\bf n}_1$, red lines) and $U_1^\dagger$ (axis ${\bf n}_2$, blue lines), the state is transformed to the appropriate $\Theta'$. The transformation is followed by the rotation $U_0^\dagger$ to achieve the target $\Phi'$.}
\label{fig:fig4}
\end{figure}

In order to gain insight into the proof we next omit the constraint $\sum_i \Delta \varphi_i = 2 \pi n$ and consider the case where rotation angles $\gamma_i$ are arbitrary. Starting with the point $(0,0,1)$ in the new coordinate system, it is easy to see that any polar angle $\Theta'$ of the rotated point in the new coordinate system can be achieved by alternately applying rotations $U_1^\dagger$ and $U_0^\dagger$. Each rotation $U_1^\dagger$ can increase $\Theta'$ by $2 \delta$, while $U_0^\dagger$ keeps $\Theta'$ constant and only rotates the point to a suitable starting point for new $U_1^\dagger$ rotation. Once the target $\Theta'$ is achieved, the rotation $U_0^\dagger$ is applied to bring the point to the target $\Phi'$ (see Fig.~\ref{fig:fig4}).

This method becomes less practical when we impose the constraint $\sum_i \Delta \varphi_i = 2 \pi n$ due to $\gamma_i$-dependent rotations. Although the unconstrained result suggests that complete coverage of the Bloch sphere may be possible, it is not clear that this remains the case when the constraint is imposed. We approach the problem numerically by generating sets of random displacements, $\Delta \varphi_i$,  which satisfy the constraint and explore the subsequent coverage on the Bloch sphere. The relevant parameters are the number of $\Delta \varphi_i$ shifts, $m$, with different $\alpha$, the number of full rotations of the electron around the ring, $n$, and the factor of amplification, $k$. By selecting a large sample of randomly transformed points, one can determine which parts of the Bloch sphere are covered for a particular choice of $m$, $n$ and $k$. We are especially interested in small values of parameters since large values of $k$ are not accessible experimentally, while large $m$ and $n$ increase the time needed to realize the transformation.

\begin{figure}[htbp]
\includegraphics[scale=1]{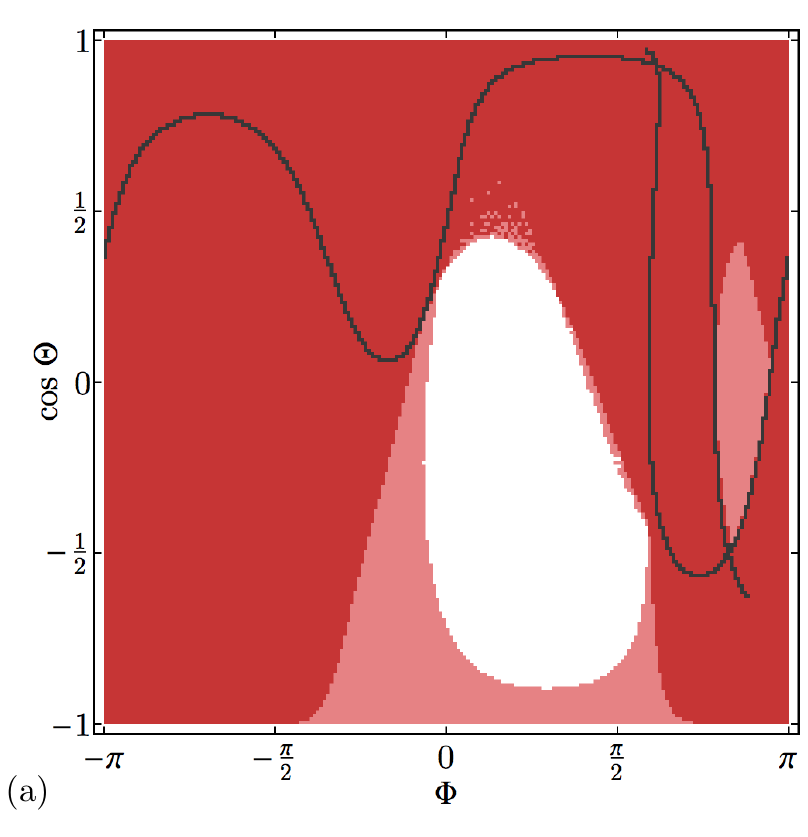} \vspace{0.cm}
\includegraphics[scale=1]{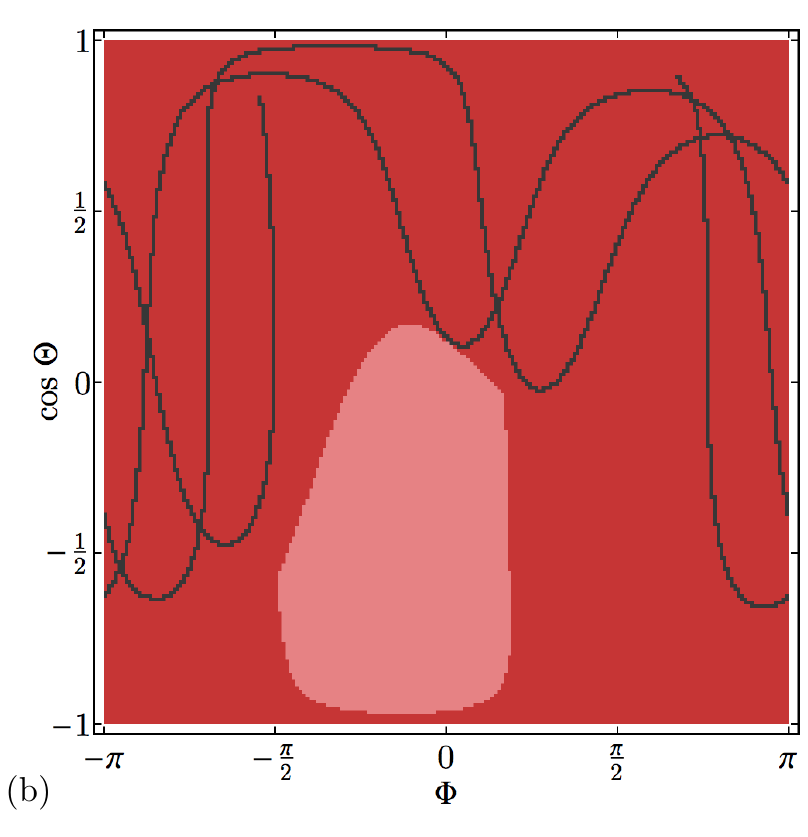}
\caption{Figure shows parts of the Bloch sphere that are coverer by $n=1$ (Fig. (a)) or $n=2$ (Fig. (b)) rotations of the electron around the ring for factor of amplification of Rashba coupling $k=5$. Black color shows the area, covered with $m=2$, dark red with $m=3$ and light red with $m=4$ shifts of electron position. For our set of $m$, the white area on Fig. (a) can only be covered for $n=2$ rotations. Calculations were performed on a $200 \times 200$ grid.}
\label{fig:fig5}
\end{figure}

The surface of the Bloch sphere, parametrized with $\cos \Theta$ and $\Phi$, was split into a $200 \times 200$ grid. For each set of parameters $m$, $n$ and $k$, a set of angles $\Delta \varphi_i$ was chosen at random, resulting in a pair of values $\left( \cos \Theta, \Phi\right)$, lying within a certain grid cell. By taking a large number ($N = 5 \times 10^6$) of sets, we ensure that each grid cell contains at least one point when the cell is in an allowed region for the selected $m$, $n$ and $k$. Thus we determine, which parts of the sphere are covered for each set of parameters, shown in Fig.~\ref{fig:fig5}.

For $m=2$, only the first shift $\Delta \varphi_1$ is a free parameter, with $\Delta \varphi_2$ being determined by constraint $\sum_i \Delta \varphi_i = 2 \pi n$. This results in points on the sphere forming a one-dimensional structure, i. e. the black line in Fig.~\ref{fig:fig5}. For $m \geq 3$, two or more shifts $\Delta \varphi_i$ are independent, resulting in two-dimensional areas on the sphere being covered.

\begin{figure}[htbp]
\includegraphics[scale=1]{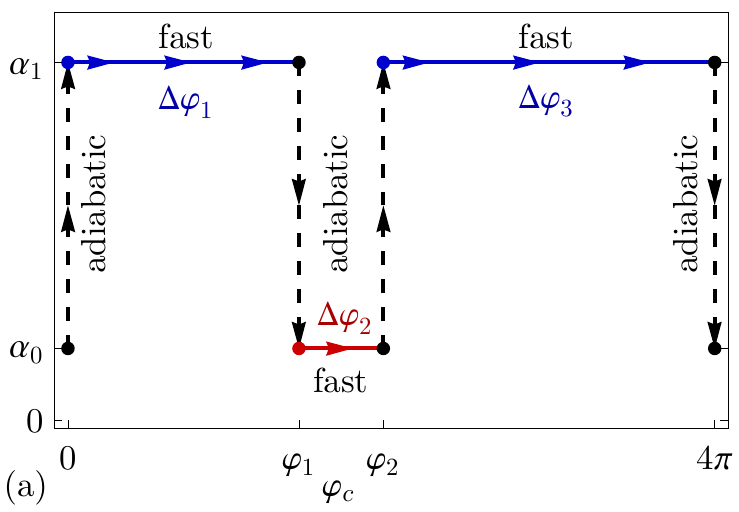}
\includegraphics[scale=1]{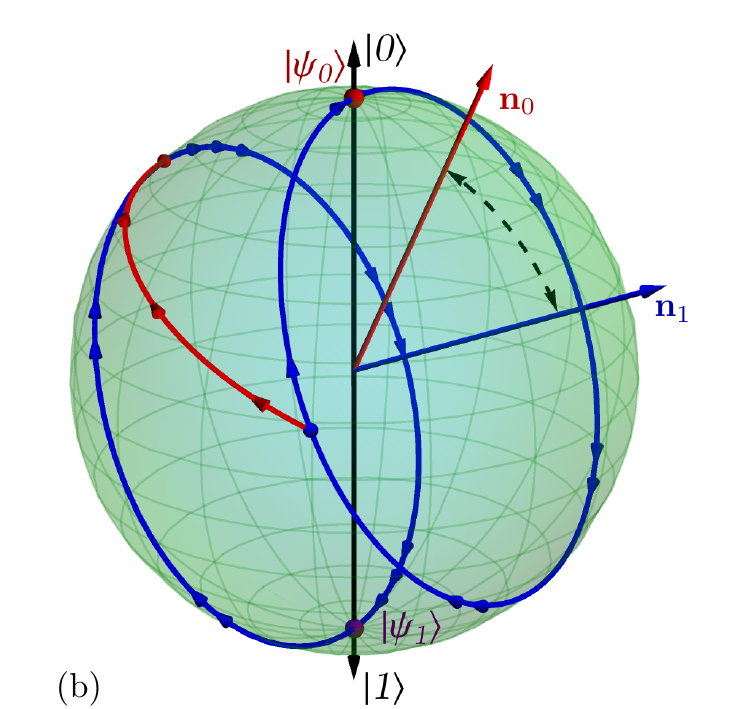}
\caption{Spin flip transformation, realized with 3 shifts of the wave function, presented in the same manner as Fig.~\ref{fig:fig3}. Shifting angles $\Delta \varphi_1=4.49$, $\Delta \varphi_2=1.64$ and $\Delta \varphi_3= 6.44$ are calculated numerically. }
\label{fig:fig6}
\end{figure}

 By keeping track of which set $\Delta \varphi_i$ resulted in a certain transformation, a table of transformation parameters was obtained for each point on the Bloch sphere, allowing us to reach each point in a controlled way. As an example, the spin flip transformation is presented in Fig.~\ref{fig:fig6}, showing the change of parameters in the $\alpha - \varphi_c$ plane and corresponding rotations on the Bloch sphere.

Following the line of thought for the omitted constraint $\sum_i \Delta \varphi_i = 2 \pi n$, we assume that $m \approx \frac{\pi}{2 \delta}$ shifts should suffice to cover the whole sphere. For $k \approx 5$ in InAs nanowires \cite{Liang2012}, this results in $m \approx 4$ shifts, which is confirmed even with the constraint in Fig.~\ref{fig:fig5}. However, for certain $n$, the average shift will be of order $\Delta \varphi_i \approx \frac{2 \pi  n}{m}$, resulting in average rotations around each axis $\gamma_i \approx \frac{2 \pi n}{m} \sqrt{1 + \alpha_i^2} $. To ensure sufficiently large rotation angles $\gamma_i \geq \pi $, assumed in the analytic explanation given earlier, the number of rotations $n$ should therefore be sufficiently large.

\begin{figure}[htbp]
\includegraphics[scale=1]{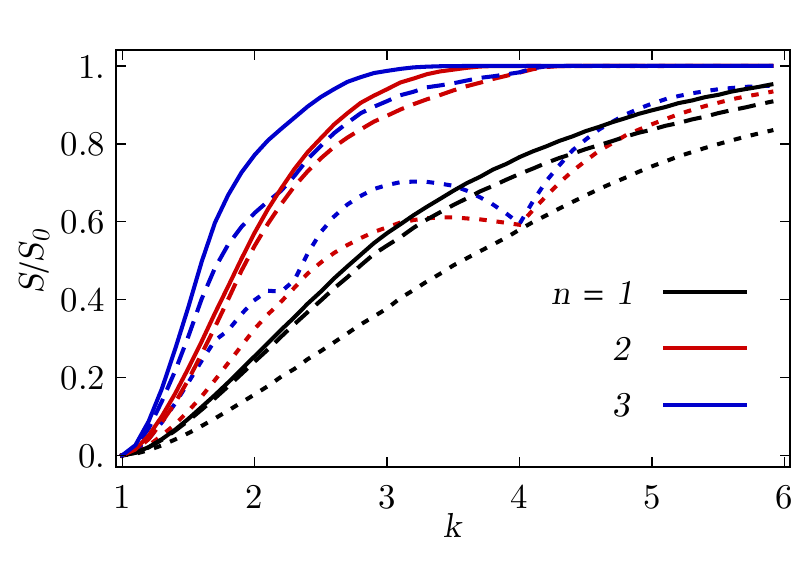}
\caption{ Accessible coverage of the Bloch sphere for certain values of $m$ and $n$ as a function of $k$ with $m=3$ (dotted), $4$ (dashed) and $5$ (solid). For $k<6$, $m=4$ and $n=2$ is the optimal choice of parameters.}
\label{fig:fig7}
\end{figure}

Since each shift $\Delta \varphi_i$ is accompanied by an adiabatic change of $\alpha$, the total time of operation will scale linearly with $m$. On the other hand, large $n$ will result in larger electron shifts, also resulting in increased time. In practice we would therefore like to keep both $m$ and $n$ as small as possible at fixed $k$. Numerically calculated coverage of the Bloch sphere for different parameters is presented in Fig.~\ref{fig:fig7} and shows that all these limitations are optimally fulfilled for $m=4$ and $n=2$.

\section{Discussion and conclusion}
\label{sec:discussion}

The effectiveness of the proposed qubit transformations in covering the Bloch sphere depends strongly on the factor of amplification of the Rashba coupling. As shown in Sec.~\ref{sec:qtrans}, the stronger the amplification, the larger the angle between axes of rotation ${\bf n}_0$ and ${\bf n}_1$, resulting in a lower number of required rotations. From this point of view, InAs nanowires with strong amplification of the Rashba coupling in the presence of an external electric field \cite{Liang2012}, are more promising candidates for experimental realisation than planar heterostructures. With the Rashba coupling $\alpha_R = 0.5 - 3 \times 10^{-2}$\,eV\,nm, the radius of the ring (Eq.~\eqref{eq:def1}) should be $R \approx 150$\,nm. Another possibility is fabricating a ring from a GaAs heterostructure for which the amplification of the Rashba parameter is smaller than in InAs ($k=1.5$ \cite{Nitta1997}), resulting in a required ring radius $R \approx 100$\,nm.

Each transformation $U_i^\dagger$ requires time $\Delta t_{\xi,i}$ to shift the wave function and time $\Delta t_{\alpha,i}$ to adiabatically change the Rashba coupling from $\alpha_0$ to $\alpha_1$ or vice versa. In a simple scheme of shifting the wave function with two instantaneous changes in $\xi(t)$, the time of the shift depends directly on the frequency of the harmonic potential $\omega$: $\Delta t_{\xi,i} = \frac{\pi}{\Omega}$ \cite{Cadez2014}. Considering the limit for the spread of the wave function $\sigma$, a realistic value for frequency is $\Omega = \frac{\epsilon}{R} \sqrt{\frac{8}{m}}$, giving for both GaAs and InAs system the estimate $\Delta t_{\xi,i} \approx 0.02$\,ns. To ensure adiabatic limit, $\Delta t_{\alpha,i}$ should be much larger than $\Delta t_{\xi,i} $. At a conservative estimate of around a hundred times this value, $\Delta t_{\alpha,i} \approx 2$\,ns, the total time needed to realize $U^\dagger$ is
\begin{equation}
t_{total} = 4 \Delta t_i \approx 10\,\text{ns}.
\end{equation}
With spin decoherence times in semiconductors of order $200\,\mu$s \cite{Bayat2013}, this has the potential to accommodate more than  $10^5$ qubit transformations.

In summary, we have shown that controlled propagation of an electron wave-packet around a semiconductor ring with tunable Rashba interaction can be used to define qubits based on a localized ground-state Kramers doublet for which qubit transformations may be performed for an integral number of revolutions. Full coverage of the Bloch sphere may be achieved provided that the Rashba spin-orbit interaction is sufficiently large and tuneable. InAs quantum wires are within the range which allows this and also has the potential for many thousand qubit operations within a spin-coherence time.

\appendix
\section{Adiabatic evolution of Kramers states}
\label{app:appA}

In this appendix we derive an expression for the time-dependent wavefunction for cases in which $\alpha(t)$ changes adiabatically at constant potential well position $\xi(t) = \xi_0$. To do this we make a specific choice for the Kramers doublet states denoted by $\psi_{\alpha(t),\xi_0,s}$ for each $\alpha(t)$ and centred around $\xi_0$, such that the adiabatic change of $\alpha(t)$ changes only the parameter $\alpha$ of each ot two states, but does not mix them. The initial Kramers state at $\alpha(t=0) = \alpha_0$, i. e. $\psi_s(\varphi,t=0) = \psi_{\alpha_0,\xi_0,s}(\varphi)$, should therefore evolve as:
\begin{equation}
\psi_s(\varphi,t) = \psi_{\alpha(t),\xi_0,s}(\varphi)
\end{equation}
These Kramers states lie in the subspace, spanned by states Eq.~\eqref{eq:psi_st0}, but with spinors $\chi_{\alpha,\xi_0,s}$ depending on $\alpha$ and $\xi_0$:
\begin{equation}
\label{eq:app_psi_alpha}
\psi_{\alpha,\xi_0,s}(\varphi) =  U^\dagger_z(\varphi) U_{\alpha}^\dagger(\varphi - \xi_0) g_0(\varphi - \xi_0) \chi_{\alpha,\xi_0,s}.
\end{equation}
We seek a solution as some spin rotation, applied to basis spinor $\chi_s$: $\chi_{\alpha,\xi_0,s} = \tilde{U}_{\alpha,\xi_0}^\dagger \chi_s$. The states will not mix if\cite{Wilczek1984}
\begin{equation}
\left\langle \frac{\partial \psi_{\alpha,\xi_0,s}}{\partial \alpha} \bigg| \psi_{\alpha,\xi_0,s'} \right\rangle = 0.
\end{equation}
for any $s$ and $s'$.
This leads to a differential equation for the rotation $\tilde{U}^\dagger_{\alpha,\xi_0}$:
\begin{equation}
\frac{\partial \tilde{U}_{\alpha,\xi_0}^\dagger}{\partial \alpha} = - \tilde{U}_{\alpha,\xi_0}^\dagger \int_{- \pi}^\pi g_0^2(\varphi) U_\alpha(\varphi) \frac{\partial U_\alpha^\dagger(\varphi)}{\partial \alpha} d \varphi.
\end{equation}
The initial condition at $\alpha = 0$ is chosen to be $\tilde{U}^\dagger_{\alpha=0,\xi_0} =\mathbb{1}$, so the states in absence of Rashba coupling will be pure spin states
\begin{equation}
\psi_{\alpha=0,\xi_0,s} = g_0(\varphi-\xi_0) \chi_s.
\end{equation}
The transformation $\tilde{U}^\dagger_{\alpha,\xi_0} $ will not depend on $\xi_0$. In the limit where the harmonic oscillator is a good approximation for the potential, the integral can be evaluated analytically by extending the limits of integration to $\pm \infty$:
\begin{equation}
\int_{- \infty}^\infty g_0^2(\varphi) U_\alpha(\varphi) \frac{\partial U_\alpha^\dagger(\varphi)}{\partial \alpha} d \varphi = \frac{i}{2} \left( \frac{e^{-\frac{\sigma^2}{4} \left(1+ \alpha ^2\right)}-1}{\left(1 + \alpha ^2\right)}\right) \sigma_y = \frac{i}{2} f(\alpha,\sigma) \sigma_y .
\end{equation}
Since this expression is proportional to the Pauli operator $\sigma_y$, the integrated result for $\tilde{U}_{\alpha,\xi_0}^\dagger$ can be expressed as a spin rotation around the $y$-axis by an angle $\tilde{\vartheta}_\alpha$:
\begin{equation}
\label{eq:app_Ualphaxi}
\tilde{U}_{\alpha,\xi_0}^\dagger = e^{-i \frac{\tilde{\vartheta}_\alpha}{2} \sigma_y} = U_y^\dagger(\tilde{\vartheta}_\alpha), \quad \tilde{\vartheta}_\alpha = \int_{0}^\alpha f(\alpha',\sigma) d\alpha'.
\end{equation}
Results for different $\sigma$ are plotted as $\vartheta_\alpha = \tilde{\vartheta}_\alpha - \arctan \alpha$ in Fig.~\ref{fig:fig2}. The angle of rotation is in fact rather small, with leading term being quadratic in $\sigma$: $\tilde{\vartheta}_\alpha\approx -\frac{\alpha \sigma^2}{4}$.

Substituting Eq.~\eqref{eq:app_Ualphaxi} into Eq.~\eqref{eq:app_psi_alpha} gives the desired Kramers doublet states:
\begin{equation}
\psi_{\alpha,\xi_0,s}(\varphi) =  g_0(\varphi - \xi_0) U^\dagger_z(\varphi) U_{\alpha}^\dagger(\varphi - \xi_0) U_y^\dagger(\tilde{\vartheta}_\alpha) \chi_s.
\end{equation}
Hence, the initial state, written in this basis:
\begin{equation}
\psi(\varphi,t=0) = \sum_s \psi_{\alpha(t=0),\xi_0,s}(\varphi) c_{0,s},
\end{equation}
will evolve adiabatically into
\begin{equation}
\psi(\varphi,t) = e^{i \phi_{\alpha}(t)} \sum_s \psi_{\alpha(t),\xi_0,s}(\varphi) c_{0,s},
\end{equation}
with an $\alpha(t)$-dependent Kramers basis which evolves while the coefficients $c_{0,s}$ remain the same. The spin independent phase factor is a time-integral of energy: $\phi_\alpha(t) = \int_{0}^t \frac{E_n(t')}{\hbar} dt' $, where time dependence of $E_n$ arises from the changing SOI.

\bibliography{ref}
\end{document}